\documentclass[useAMS,usenatbib]{mn2e}
\usepackage{epsfig}

\def\lcdm{$\Lambda$CDM }

\newcommand{\hMpc}{\mbox{$h^{-1}$ Mpc} }

\newcommand{\hMpcCom}{\mbox{$h^{-1}$ Mpc,} }
\newcommand{\hMpcDot}{\mbox{$h^{-1}$ Mpc.} }
\newcommand{\hMpcDotK}{\mbox{$h^{-1}$ Mpc.)} }

\newcommand{\hmsun}{\mbox{$h^{-1}$ $M_{\odot}$} }

\newcommand{\hmsunCom}{\mbox{$h^{-1}$ $M_{\odot}$,} }
\newcommand{\hmsunDot}{\mbox{$h^{-1}$ $M_{\odot}$.} }

\newcommand{\kms}{\mbox{km s$^{-1}$} }

\newcommand{\kmsDot}{\mbox{km s$^{-1}$.} }

\newcommand{\MpcInv}{\mbox{$\tx{Mpc}^{-1}$} }

\def\la{\mathrel{\hbox{\rlap{\hbox{\lower4pt\hbox{$\sim$}}}\hbox{$<$}}}}
\def\ga{\mathrel{\hbox{\rlap{\hbox{\lower4pt\hbox{$\sim$}}}\hbox{$>$}}}}

\def\gsim{\ga}
\def\lsim{\la}

\newcommand{\bc}{\begin{center}}
\newcommand{\ec}{\end{center}}

\newcommand{\be}{\begin{equation}}
\newcommand{\ee}{\end{equation}}

\newcommand{\tx}[1] {\rmn{#1}}

\title{The case for non-gaussianity on cluster scales}
\author[H. Mathis, J.~M.~Diego and Joseph Silk]
	{H.~Mathis$^1$\thanks{Email: hxm@astro.ox.ac.uk}, 
	J.~M.~Diego$^2$ and J. Silk$^1$
	\\
        $^1$Department of Astrophysics, University of Oxford, Keble Road, OX1 3RH, UK \\
        $^2$Department of Physics and Astronomy, UPenn, 209 South 33rd St, Philadelphia PA 19104, USA }
	
\date{Draft version \today}

\begin{document}

\maketitle

\label{firstpage}
 

\begin{abstract}
We address whether possible scale-dependent deviations from
gaussianity in the primordial density field that are consistent with the
cosmic microwave background  observations  could explain
the apparent excess of early cluster formation at high redshift.  Using two
phenomenological non-gaussian models we find that at fixed
normalisation to the observed local abundance of massive clusters, the
protoclusters observed at z$\sim$4 are significantly more likely to
develop in strongly non-gaussian models than in the gaussian
paradigm. We compute the relative $z<1$ evolution of X-ray cluster
counts in the non-gaussian case with respect to the gaussian
expectation, and the relative excess contribution to the CMB power
spectrum due to the integrated thermal SZ effect.  We find that both
the observed hints of an  unexpectedly slow evolution in
 the X-ray counts and the excess power
at high $\ell$ that may have been
observed by CMB interferometers  can also be reproduced in 
our non-gaussian simulations.
\end{abstract}




\section[]{Introduction}
\label{sec:Intro}
 	
Non-gaussianities in the primordial perturbations of the cosmic density field 
are inherently difficult to generate in simple versions of inflation, and 
most inflationary models predict negligible amounts of relic non-gaussianity.  
Multiple field inflation or more speculative models such as those assuming 
non-homogeneous reheating processes or those seeded with topological defects 
 can however readily rise to the challenge  of  spawning deviations from a 
gaussian initial density field.  Of course, the amount of allowed primordial non-gaussianity 
on CMB scales has come under increasing pressure from the recent results 
of \emph{WMAP} \citep{Kom03}, and constraints are expected to tighten with upcoming data releases.

However current constraints from all-sky CMB maps only transfer to 
constraints at cluster or galaxy scales in the case of scale-invariant
non-gaussianity.  If non-gaussianity is strongly scale-dependent and 
arises only at cluster scales, it can simultaneously fit  the
\emph{WMAP} bounds yet affect large-scale structure so as to
induce predictable deviations from the gaussian paradigm. The caveat here   
is  the  ``fine-tuning'' of the physical scale at which 
non-gaussianity appears, which has to fall between $\ell\sim300$ and 
$\ell\sim$ 2500. (Models including defects such  
as linear cosmic strings as additional seeds for structure formation 
would naturally have such a scale, see \citealt{Ave98}). Our 
approach here is phenomenological, 
and  we consider  two hypothetical models which have such 
scale-dependent non-gaussianity, and, incidentally, are also to some 
extent motivated by theoretical considerations.

From the point of view of structure formation at low redshift and
relatively small scales, issues for the gaussian \lcdm paradigm
include the early formation of protoclusters up to $z\sim4$ as 
 traced by radio galaxies surrounded by Ly-$\alpha$ emitters
and Lyman-break galaxies, the
observed apparently slow evolution of the comoving density of bright X-ray
clusters, and the possible excess of angular power in CMB temperature
fluctuations seen by CBI and ACBAR on cluster scales that most likely
is  due to the thermal SZ effect.  It is well known that  
non-gaussianity may be able
to naturally explain possible  features associated with 
an excess (relative to Gaussian predictions) in the frequency at high
redshift of galaxy  protoclusters and
clusters, by reasoning as follows.  Rare peaks in the primordial density
field define the locations where massive clusters will form, and the
higher the peak the earlier the collapse to a virialized system. If
the probability density function (PDF) of the density field shows an
excess of high peaks compared to the gaussian case, (proto-)clusters
of given mass form earlier, and systems above a given mass will be
more abundant at fixed redshift.  Note that more speculative topics
such as  the periodicity noted by \citet{Broad90}, the presence of large
regions underdense in galaxies \citet{Fri03} or the high correlation
length of very massive X-ray clusters could also hint at
primordial non-gaussianity. Note also that scale-dependent 
non-gaussianity has been advocated by \citet{Ave04} and \citet{Che03} as a possible way 
to reconcile the high optical depth to reionization with realistic
efficiencies for the formation of the first stars, and that 
its possible impact on the inner profiles of dark matter haloes has 
been addressed by \citet{Av03}.

Here, we use simulations of large-scale structure to check the impact
of strongly positively and negatively skewed non-gaussianity on the
above observables. We evolve two primordial cosmic density fields with
a scale-dependent non-gaussianity which is at least of the same order as
the gaussian component at the scales probed by the simulation box. We
employ the concordance cosmological parameters and normalise the
models so that they give a similar abundance of massive clusters at
$z=0$. The two non-gaussian models differ by the sign of their
primordial skewness: we choose a model with primordial voids
\citep{Gri03} to realize negative skewness, and a model with a
chi-square PDF of the density field to have positive skewness (this
case been worked out in the scale-independent case by
\citealt{Pee99a}).  We compare to the gaussian \lcdm case they bracket
from the point of view of cluster formation.

The paper is organised as follows. In Section~\ref{sec:ObsHints} we
review recent observational hints of  non-gaussianity on cluster
scales.  Section~\ref{sec:Simus}  describes the setup of the
simulations and compares to observations. We conclude in
Section~\ref{sec:CCL}.



\section[]{Indicative observations}
\label{sec:ObsHints}

	We describe here the three observational hints which we can
	probe using simulations without recourse to any model for galaxy
	formation.

\subsection{High redshift assembling structures}
\label{sec:ObsHints:ProtoClusters}

\citet{Wi00} was among the first to set constraints on  possible
primordial non-gaussianity using the presence of high-redshift massive
systems. Even if some of the mass estimates he used for the $z=0.83$
cluster MS 1054-03 were probably biased (\citealt{Hoe00,Neu00})
because the system might not be dynamically relaxed, the precedent was
set.

On the theoretical side, \citet{RGS00} (see also
\citealt{Koy99,RB00,Mat00}) have shown that given
$\Omega_{\rmn{matter},0}$, the power spectrum and its normalisation to
the cluster abundance today, the mere presence of massive high
redshift clusters provided useful constraints on relative amounts of
generic non-gaussian contributions to the primordial density field.

More recently, \citet{Mil04} (see also \citealt{Kur03}) report the
detection of 21 Ly-$\alpha$ emitters assembling around a radio galaxy
at $z\sim4.1$. The full structure has a proper spatial extent of 3 to
5 Mpc. Even if the system is not virialized, \citet{Mil04} could
compute its formal velocity dispersion from the Ly-$\alpha$ systems
and found it to be of order 325 \kmsDot The enclosed mass is comparable
to that of local galaxy clusters. At $z\sim2.1$, \citet{Kur03} 
find a close pair of two such Ly-$\alpha$ systems; when 
they are considered seperately, they have velocity dispersions of, 
respectively, $\sim200$ and $\sim400$ \kmsDot
When they are considered a single group, the estimated 
velocity dispersion is 1000 \kmsDot

\subsection{Evolution of the X-ray cluster luminosity function}
\label{sec:ObsHints:XClusters}

The X-ray window is ideal for studying the evolution of massive clusters
because the selection function effects can be much better understood
than in the optical. No direct mass measurement is needed and one can
use the observed relations between the X-ray luminosity and
temperature ($L-T$) and between the temperature and mass ($M-T$) to
select massive members from a catalogue of X-ray clusters. Inevitably
however, results on the evolution of the abundance of X-ray clusters
will depend on the adopted scaling relations.

Evolution of the abundance of X-ray luminous clusters has become one
of the major tools for constraining the present day cosmological matter density
$\Omega_{\rmn{matter},0}=1$ (for instance, \citealt{Sad98}; see \citealt{Bor01} 
for evidence for a low-density universe). However, at fixed cosmological parameters, 
fixed scaling relations and assuming  normalisation to the local cluster abundance, increasing the amount of
skewness in the primordial density field changes the $z\lsim1$
evolution in the abundance of massive clusters.

Deep \emph{ROSAT}, \emph{Chandra} and \emph{XMM} X-ray observations
have revealed a significant number of distant luminous clusters
\citep{Ros00}.  At redshifts $z\lsim0.8$, statistics of X-ray clusters
have become sufficiently good  to provide realistic estimates of their
cosmological abundance. \citet{Vik03,Voe04} derive cosmological
constraints from the evolution of the cluster baryon mass fraction
which they relate to the total cluster mass, thereby skipping the
usual uncertainties in the scaling relations. Using the 160 deg$^2$
\emph{ROSAT} survey, they obtain results in line with the evolution of
the mass function expected in a concordance gaussian CDM cosmology as
they find moderately negative evolution of the cluster abundance to
$z\sim0.6$.

Recently, \citealt{Vau03} prefer an $\Omega_{\rmn{matter},0}=1$
universe from the reduction of the same data, but caution that a
variation in the scaling of the $M-T$ relation with redshift might
bias the analysis. In this work we will rederive the
evolution of the density of luminous clusters from the
$\sim$200-member cluster catalogue of \citet{Vik98} (updated using
spectroscopic redshifts in \citealt{Mul02}; we will refer to this
catalog as $VM$) and we will compare it with predictions from gaussian
and non-gaussian models, at fixed scaling relations.

\subsection{Small-scale CMB power spectrum}
\label{sec:ObsHints:CMB}

\citet{Maj00} used the upper bounds on the rms CMB  temperature fluctuations observed 
at arcmin scales by ATCA as constraints for structure formation models. They ruled out 
the possibility of a high $\Omega_{\rmn{matter},0}$  assuming \emph{COBE}  
normalisation  and gaussian fluctuations of the primordial density field.  More recently, 
high resolution CMB experiments such as CBI \citep{Mas03} or ACBAR
\citep{Kuo04} find an excess of small-scale ($\ell \approx 3000$ or 
$\theta \approx$ few arcmin) power with respect to the expectations from 
primary CMB fluctuations.

Non-gaussian models deviate more from gaussian models at high redshift
(at fixed present day normalisation). Because the SZ temperature
decrement of the CMB is independent of redshift, the angular power
spectrum of the thermal SZ effect is sensitive to the high redshift
abundance of clusters. This turns the integrated SZ power spectrum into
another efficient means of distinguishing gaussian from non-gaussian
models, in addition to the evolution of the abundance of massive
clusters.

Given the known local abundance of clusters, gaussian models
require low values of the normalisation ($\sigma_8 \approx 0.8$ if
$\Omega _{\rmn{matter},0}=0.3$) when they are fitted to the data. 
It is not clear yet if the observed excess is due to systematic
effects or if it is a real signature coming from compact sources
(point sources and/or galaxy clusters). In case this excess in the CMB
power spectrum at $\ell > 3000$ is confirmed by future experiments to
be due to galaxy clusters, it would imply a high normalisation
($\sigma_8 > 1$ for $\Omega _{\rmn{matter},0} \approx 0.3$) in the
gaussian case. Alternatively, it would point towards a combination of
the $M-T$ and $r_{\tx{core}}-M$ relations (detailed in
paragraph~\ref{sec:Simus:Compare:CMB} below) which would be rather
extreme (see for example \citealt{Maj01}).  A more natural explanation
however is an over-abundant population of clusters at high redshift,
in turn a possible consequence of non-gaussianity of the primordial
density field. We will consider this possibility in line with the
evolution of the scaling relations that we assume to compute the X-ray
cluster evolution.


\section[]{The case for non-gaussianity}
\label{sec:Simus}

We first describe the simulations and characterize the models with
simple statistics.  Then, we directly measure in the simulations the
three observables of the previous Section.
 
\subsection{Simulations setup} 
\label{sec:Simus:Setup}

We consider two non-gaussian models: the first, called $V$ is a model
with compensated primordial voids \citep{Gri03,Ma03a}, the second
(hereafter model $C^2$) is a model where the PDF of the primordial
field density is the square of a gaussian.  In theory the models  could be
associated with, respectively, a possible first-order transition
occurring during inflation, and to versions of multiple scalar field
inflation with isocurvature-type initial perturbations in a massive
scalar field playing the role of CDM (such an ICDM model was worked
out in detail by \citealt{Pee99a}). While the first-order phase
transition model yielding primordial voids would be truly gaussian on
CMB scales, this is not the case for the ICDM model.  Tuning the ICDM
model initially suggested so that it has the correct scale-dependent
non-gaussianity is beyond the scope of this paper. However 
in conformity with our phenomenological 
 approach, we will refrain in the following from linking 
these $V$ and $C^2$ realisations of the initial density field to  a
specific  inflationary scenario. Rather, we simply take them as
possible representations of the primordial density field on cluster
scales that could arise in generic inflationary models.
 For reference, we also consider a gaussian model $G$.

\begin{figure}
\begin{minipage}{9cm}
\epsfig{file=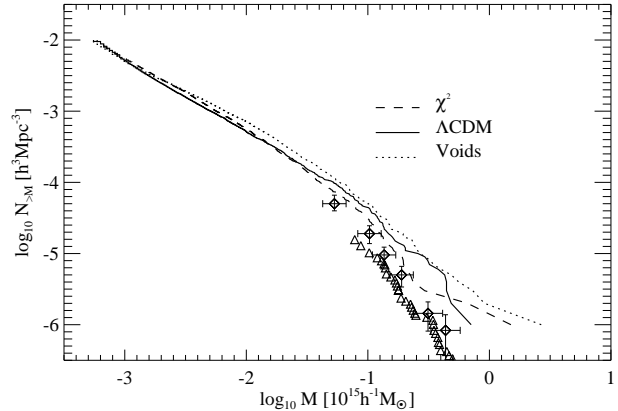,width=9cm}
\caption{Mass function of dark matter haloes at $z=0$ for our three
phenomenological models.  Diamonds with error bars (resp. triangles)
are the data of \citet{Bah03a} \citep{Vik03}.}
\label{fig:MF}
\end{minipage}
\end{figure}

All three models share the same \lcdm cosmological parameters
$\Omega_{\rmn{CDM},0}=0.3$, $\Omega_{\Lambda,0}=0.7$ and $h=0.7$.  We
impose the same initial matter power spectrum in the $V$ and $G$
models: $n_{\rmn{s}}$=1 with an adiabatic CDM transfer function. In
the $C^2$ model, the initial power spectrum is that used by
\citealt{Pee99a} with $n_{\rmn{s}}$=-1.8 with an isocurvature CDM
transfer function.  We ensure that the $z=0$ cluster mass functions of
the non-gaussian models match reasonably well those of the gaussian
\lcdm model with $\sigma_{8}=0.9$. This is shown in 
Fig.~\ref{fig:MF}, where we compare the mass functions in our three
simulated models to the local optical cluster mass function of
\citet{Bah03a} and to the X-ray cluster mass function of
\citet{Vik03}.  Differences with respect to the gaussian case are
limited to 25 percent up to $1.6\times10^{14}$\hmsun cluster
masses. At masses $\sim10^{15}$\hmsunCom differences reflect
cosmic scatter.  In practice, the match to the \lcdm $G$ model was
achieved in the $C^2$ model by normalising the initial power computed
from the realized field so that linear extrapolation yields
$\sigma_{8}=0.65$ today. In $V$, the underlying gaussian field before
``addition'' of the voids has a normalisation similar to that of
$G$. Note that the parameters of the distribution of void sizes in the
$V$ model was set so as to match at present day the large voids seen
in local galaxy surveys \citep{ElAd00,Hoy03}. As such, $V$ would also
be a natural solution to the presence of large regions in the nearby
Universe which are completely empty of bright galaxies \citep{Pee01}.

\begin{figure}
\begin{minipage}{9cm}
\epsfig{file=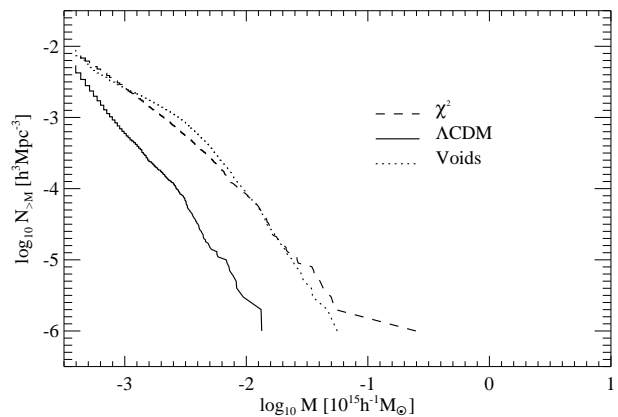,width=9cm}
\caption{Mass function of dark matter haloes at $z=4$ for our three
phenomenological models.}
\label{fig:MFz4}
\end{minipage}
\end{figure}

The simulations start at high redshift $z_{\tx{init}}=300$ and follow
collisionless structure formation using 128$^3$ particles in a 100
\hMpc box with periodic boundary conditions. We employ the publically
available code N-body code {\sc GADGET} without hydrodynamics
\footnote{\tt
http://www.mpa-garching.mpg.de/$\sim$volker/gadget/index.html} and set
the equivalent Plummer softening length to one tenth of the mean
interparticle separation. The mass of a dark matter particle is
$3.94\times10^{10}$ \hmsunDot

\subsection{Non-gaussian signatures in the simulations} 
\label{sec:Simus:NonGaussSig}

\begin{figure}
\begin{minipage}{9cm}
\epsfig{file=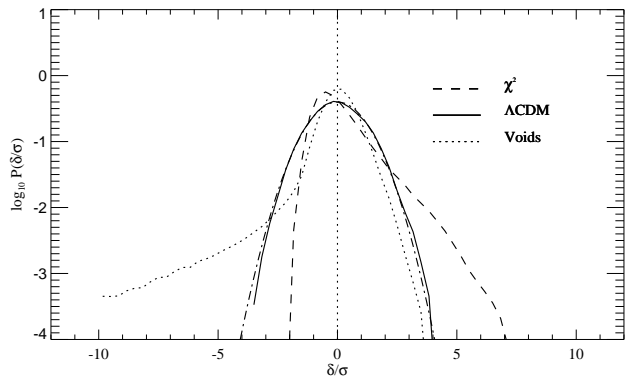,width=9cm}
\caption{PDF of the rescaled overdensity field $\delta/\sigma$
measured at $z=50$ in the simulations, for the two non-gaussian and
for the gaussian model. A gaussian distribution (dash-dotted line) has
been overplotted.  Note the opposite skewness of the $C^2$ and $V$
models.}
\label{fig:PDF}
\end{minipage}
\end{figure}

This specific choice of the strongly non-gaussian models $V$ and $C^2$
is attractive in the sense that the skewness of the primordial
density field approximately brackets that of a gaussian field.  This
is illustrated in Fig.~\ref{fig:PDF} where we plot the PDF of the
rescaled overdensity field $\delta/\sigma$ directly measured it in the
three simulations at $z=50$ over 8 \hMpc scales.  Dotted, solid and
dashed lines respectively correspond to the $V$, $G$ and $C^2$
cases. For comparison, a gaussian has been overplotted in dash-dotted
line. While the PDF of the density of the $V$ simulation is strongly
negatively skewed due to the presence of the voids, that of the $C^2$
simulation is positively skewed as expected. (Note that the $C^2$ and
$G$ simulations are still within the ``linear regime'' over 8 \hMpc
scales by $z=50$, while the expanding primordial voids in the $V$
simulation have already fully emptied regions as large as 8 \hMpcDotK

Such skewed primordial PDFs directly impact the $z=4$ halo mass
function as shown in Fig.~\ref{fig:MFz4}. There, we find early
structure formation to be a common feature of the $C^2$ and $V$
models. In the $V$ model, this is the result of two effects:
fragmentation of the shell surrounding the voids, and non-linear
motions in the background which are induced by the void
underdensities. In the $C^2$ model, it is the effect of the relative
primordial overabundance of $>3\sigma$ peaks with respect to the
gaussian case.

\begin{figure}
\begin{minipage}{9cm}
\epsfig{file=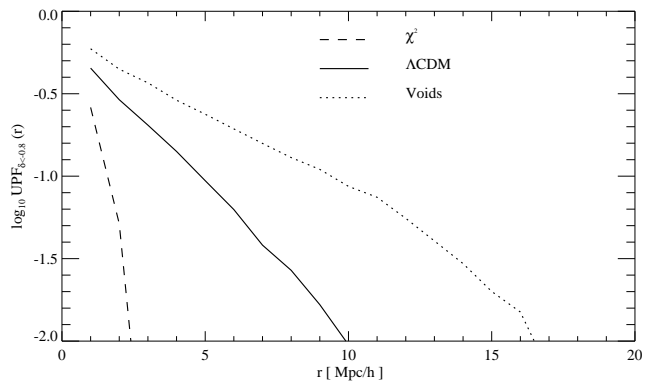,width=9cm}
\caption{Underdensity probability function $\rmn{UPF}_{<\delta}$(r)
for $\delta=-0.8$ measured at $z=0$ in our three models.  Note the
differences in the non-gaussian models $V$ and $C^2$ (dotted and
dashed lines) compared to the gaussian case (solid line). The void
model naturally has a higher UPF(r), while the $C^2$ model has very
few underdense regions larger than 1 \hMpc in radius.}
\label{fig:UPF}
\end{minipage}
\end{figure}

The effect of the three different initial conditions is
also clearly mirrored, although differently, in the underdensity probability function
(UPF) $\rmn{UPF}_{<\delta}$(r) \citep{Wein92}.  We define the UPF as
the probability that the matter underdensity $\delta$ in a randomly
placed sphere of comoving radius $r$ has a value less that -0.8,
following \citet{Hoy03}. Fig.~\ref{fig:UPF} shows the underdensity
probability function at $z=0$ for the gaussian $G$, the $C^2$ and the
$V$ models (solid, dashed and dotted lines respectively).  Obviously,
the void model has a larger probablility of underdensities compared to
the gaussian case, while the $C^2$ model has very few regions with a
radius larger than 1 \hMpc where $\delta\leq-0.8$, a consequence of the
$\chi^2$ distribution of the PDF of the primordial field.  A
comparison to results from local galaxy surveys is nevertheless out of
the present scope: this would require a model for galaxy bias, which plays a
critical role in void statistics.

\subsection{Comparison to observations} 
\label{sec:Simus:Compare}

We first compute statistics relevant to the protoclusters observed at
$z\sim4$.  Then, we choose two models for the cluster scaling
relations and employ them to compare the observed evolution of the
X-ray luminosity function.  Finally, we estimate the expected thermal
SZ angular power on cluster scales.

\subsubsection{Finding high-redshift protoclusters} 
\label{sec:Simus:Compare:ProtoClusters}

Comparing the ``protocluster'' observations of \citet{Mil04} to
simulations of structure formation requires suitable statistics. We
proceed as follows. At $z=4$, we randomly throw a large number (30000)
of spheres of proper radius $r_{\tx{throw}}$ in the simulation and
compute for each the  overdensity with respect to the background and
the one-dimensional velocity dispersion $\sigma_{v}$ of all the  dark matter particles
with respect to the  centre of mass. We then construct the PDF of
$\sigma_{v}$ gathering all the spheres of adequate overdensity. (This
procedure was suggested by \citealt{RB00} in the case of the sampling
of a density field.)  Note that this method does not require  
the overdensities within the spheres to have virialized.

\begin{figure*}
\begin{minipage}{18cm}
\epsfig{file=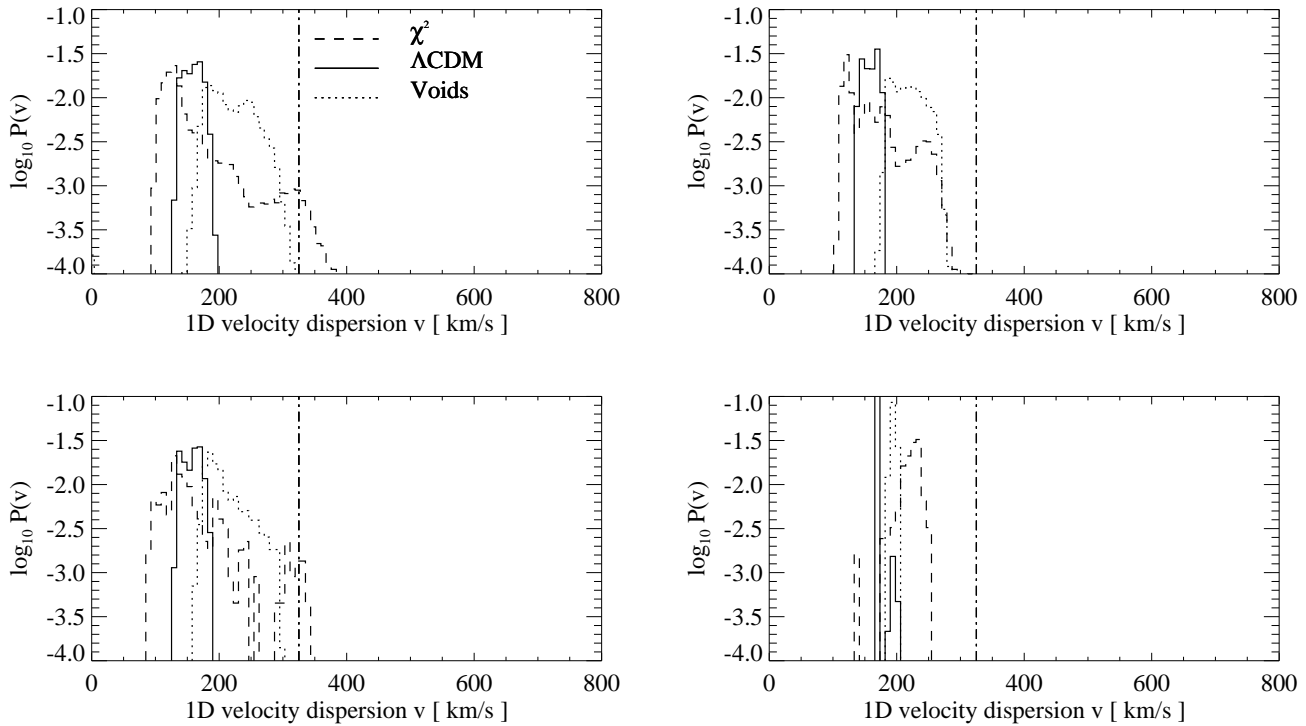,width=18cm}
\caption{$z=4$ PDF of the inner one-dimensional velocity dispersion measured from a 
large number of spheres thrown in the simulations. $r_{\tx{throw}}=5$
\hMpc is on the left, $r_{\tx{throw}}=9$ \hMpc is on the right. Upper
row: spheres in all environments, lower row: spheres where $\delta \in
[1,3]$.  Dotted, solid and dashed lines are for the $V$, $G$ and $C^2$
simulations respectively. The vertical dash-dotted line is the system 
measured by \citet{Mil04} at $z=4.1$.}
\label{fig:ThrowPDFz4}
\end{minipage}
\end{figure*}

\begin{figure*}
\begin{minipage}{18cm}
\epsfig{file=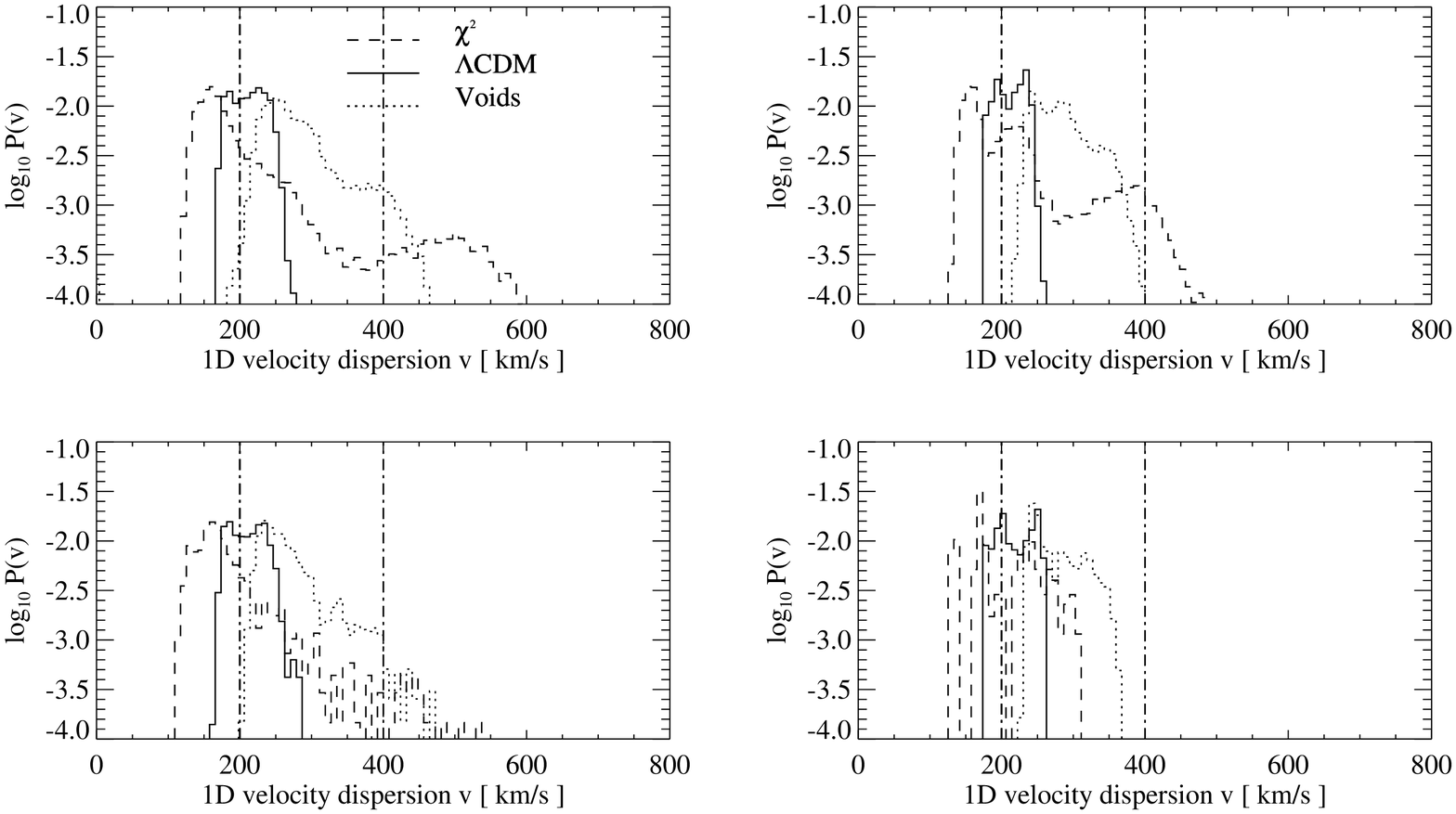,width=18cm}
\caption{Same as Fig.~\ref{fig:ThrowPDFz2} but at z=2. Vertical 
dash-dotted lines correspond to the $\sim200$ and 400  \kms 
velocity dispersion estimates of \citet{Kur03}.} 
\label{fig:ThrowPDFz2}
\end{minipage}
\end{figure*}

Fig.~\ref{fig:ThrowPDFz4} shows the resulting PDF at $z=4$. The upper
left panel corresponds to throwing spheres with $r_{\tx{throw}}=5$
\hMpc comoving and gathers the PDF of $\sigma_{v}$ measured in the
spheres irrespective of their environment (overdensity with respect to
the background) . The 3 to 5 Mpc proper extent reported by 
\citealt{Mil04} maps on to a 5 to 9 \hMpc comoving sphere radius. 
 The upper right panel of the same figure
corresponds to throwing spheres with $r_{\tx{throw}}=9$ \hMpc again
selected in all environments. The lower row of
Fig.~\ref{fig:ThrowPDFz4} is an addition to  the upper row, for spheres
whose dark matter overdensity $\delta_{\tx{DM}} \in [1,3]$.  Dotted,
solid and dashed lines are for the $V$, $G$ and $C^2$ simulations
respectively.  The \citet{Mil04} 325 \kms velocity dispersion measure
for the system of 21 Ly-$\alpha$ emitters at $z=4.1$ is shown as
the vertical dash-dotted line. Fig.~\ref{fig:ThrowPDFz2} repeats the
same statistics at $z\sim2$, where other such protoclusters have been
observed. The 200 and 400 \kms velocity dispersion 
estimates of \citet{Kur03} are shown with vertical dash-dotted lines. 
(For the purpose of comparison we have kept the 5 and 9 \hMpc comoving 
radii at $z=2$.)

It is clear from these figures that at $z=4$, there is no 5 \hMpc
comoving radius sphere in the gaussian simulation $G$  whose inner
velocity dispersion is within 100 \kms of what is observed, even
regardless of environment.  The situation is much more favourable for
the other two non-gaussian models which we explore, where a non-zero
PDF overlaps with the observations.  The hypothesis here is clearly the
absence of velocity bias between the simulated dark matter and the
Ly-$\alpha$ emitters. For $r_{\tx{throw}}=9$ \hMpcCom none of our
simulations matches the observations.

\subsubsection{Simulating the abundance of luminous X-ray clusters} 
\label{sec:Simus:Compare:XClusters}

\begin{figure}
\begin{minipage}{9cm}
\epsfig{file=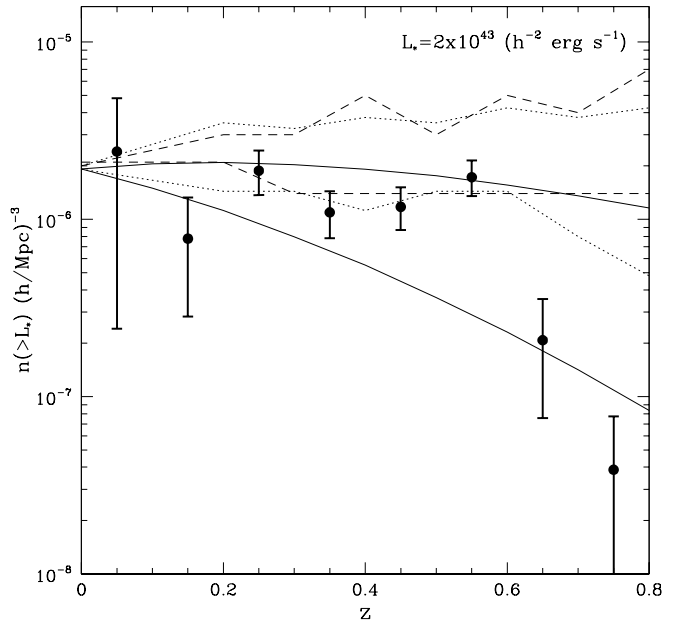,width=9cm}
\caption{ Evolution of the density of luminous clusters with redshift.
            The sample of clusters was taken from the \citet{Vik98}
            and \citet{Mul02} catalog.  Dotted and dashed lines
            show the expected evolution for non-gaussian $V$ and $C^2$
            simulations respectively, while solid lines
            correspond to the gaussian case. In both cases, the upper
            line corresponds to a model where the $L-M$ relation
            evolves as $L \propto M^2(1+z)^3$ while the lower line is
            for a model with no evolution ($L \propto M^2$). }
\label{fig:NL_z}
\end{minipage}
\end{figure}

We first select the most luminous clusters ($L_{\rmn{X}} > L_*$) from
the $VM$ catalog and we build their density $n(> L_*) $ as a function
of redshift:
\begin{equation}
n(> L_*) = \sum_i \frac{1}{V_i}
\end{equation}
where $L_*$ is the threshold luminosity and the sum is over all the
clusters above $L_*$ in the redshift bin $i$ (we take $\Delta z =
0.1$).  $V_i$ is the \emph{search} volume defined by the product of
the volume in the shell of $\Delta z = 0.1$ at redshift $z$ times the
effective sky coverage $f(L,z)$.  For the 160 deg$^2$ \emph{ROSAT}
survey used here, $f(L,z)$ was computed in \citet{Vik98}.  In
figure~\ref{fig:NL_z} we show results for a threshold $L_* = 2 \times
10^{43} h^{-2}$ erg s$^{-1}$ cm$^{-2}$.  Note that because we employ
$H_0 = 100 h$ \kms \MpcInv throughout, the original luminosities given
for $h=0.5$ in the $VM$ catalog have to be rescaled. For example,
$2\times 10^{43}\ h^{-2}$ erg s$^{-1}$ cm$^{-2}$ correspond to an
original $0.8 \times 10^{44}$ erg s$^{-1}$ cm$^{-2}$ for $h=0.5$).

Fig.~\ref{fig:NL_z} compares the observed $n(> L_*)$ with our
non-gaussian simulations (dotted and dashed lines are for $V$ and
$C^2$ respectively) and the gaussian $G$ simulation (solid 
lines). To make proper comparison between the models, the non-gaussian
abundances have been slightly rescaled to have exactly the same $z=0$
abundance of $L_*$ clusters as in the $G$ simulation.  To take into
account the uncertainties in the cluster scaling relations we have
compared two different assumptions on the evolution of the scaling
relations.  Lower and upper lines are respectively for a no evolution
model $L \propto M^2$ and for a model where $L \propto M^2(1+z)^3$.
The gaussian model could be compatible with the observed density of
luminous clusters which is flat at least up to $z=0.6$ only if there
is a strong evolution in the $L-M$ relation up to this redshift. 
It is interesting to compare our Fig.~\ref{fig:NL_z} with figure 9 in \citet{Mul04}.
\citet{Mul04} present a curve similar to ours also using the 160
deg$^2$ sample, but they choose a higher $L_{*}=10^{45}
h_{50}^{-2}$ erg s$^{-1}$.  They similarly find a decrease in the
density of luminous clusters at $z>0.6$.  When they fit the observed
density with an evolving Schechter luminosity function normalised to
match the $z=0.2$ observed luminosity function of X-ray clusters, they
recover this trend at $z>0.6$. Their result is independent of any 
assumption (such as gaussianity) on the evolution of the mass 
function. In the non-gaussian case the $L-M$ would  
need not to evolve or even to evolve negatively to explain the $z=0.6$ drop. 
Whether this high-redshift drop is real or is due to the lack of statistics 
is an issue which has to be addressed with deeper and wider 
surveys.

In fact, the $L-M$ relation has been poorly determined so far, but we
can use the recent attempts made to constrain the evolution of the
$L-T$ and $M-T$ relation to constrain the evolution of the $L-M$
relation. \citet{Vik02b} constrain the evolution of the $L-T$ relation
as $L \propto T^{\alpha}(1+z)^{\gamma}$ with $\alpha = 2.64$ and
$\gamma = 1.5 \pm 0.3$.  \citet{Hold02} find a softer value of $\gamma
= 0.3 \pm 0.2$.  The BMW cluster survey \citep{Mor01} found evidence
for a non-evolving $L-T$ relation (or even a negative evolution).

For the $M-T$ relation, \citet{Vik03} find $T \propto M^{\chi}
(1+z)^{\psi}$ with $\psi = 0.5 \pm 0.4$.  The spherical collapse
model predicts $\psi = 1$ for $\Omega _{\rmn{matter},0} = 1$.
Combining the previous constraints one can derive that $L \propto
M^{\alpha \chi} (1+z)^{2-3}$.

\citet{Ros02} gave  a comprehensive review of the situation and suggest
 that a weak evolution in the $L-T$ relation ($\gamma < 1$) is a
 reasonable assumption.  If both the $L-T$ and the $M-T$ relation are
 non-evolving then the $L-M$ relation must also be a non-evolving
 relation. With the current uncertainties in the observed cluster
 scaling relations we stress that it is unclear if the observed
 flatness in $n(> L_*)$ is due to an evolving $L-M$ or if it is due to
 an overdensity of high redshift clusters (when compared with gaussian
 models).  Again, our purpose here is mainly to bring out the
 differences between gaussian and non-gaussian models at fixed scaling
 relations.

We can use the following simple arguments to infer the redshift dependence
of the $L-M$ relation, if we assume the validity of a theoretical mass
function model and place ourselves in the gaussian case. We use 
the fact that the luminosity function is compatible with no evolution
up to redshift $z \approx1$ (for instance figure 9 in
\citealt{Ros02}).

We simulated the luminosity function at different redshifts for a flat
cosmological model ($\sigma _8 \sim 0.9$, $\Omega _{\rmn{matter},0} =
0.3$).  The luminosity function can be derived from the mass function
through the relation:
\begin{equation}
\frac{dn}{dL} = \frac{dn}{dM}\frac{dM}{dL}.
\end{equation} 
Given the mass function (for example from the Press-Schechter
formalism or its extensions to the non-gaussian cases), the luminosity
function only depends in the assumed $L-M$ relation.  Requiring that
the luminosity function must be similar at low, intermediate and high
redshifts we can constrain the redshift dependence of the $L-M$
relation.  In a simple test we have assumed the form:
\begin{equation}
L = L_{0} M^{\beta} (1+z)^{\zeta},
\end{equation}
 changing $\beta$ and $\zeta$ and looking for the combination which
yields a similar luminosity function at different redshifts ($\zeta$
does not show a particular dependence with $L_0$).  We have found that
only models close to the relation $\zeta = 1.5\beta - 1.1$ produce a
non-evolving luminosity function.  If $\beta$ is derived from the
observed $T \propto M^{0.55-0.66}$ and $L \propto T^{2.5-3}$ then
$\beta \approx 1.4-2$.  The corresponding $\zeta$ for a non-evolving
luminosity function reads $\zeta \approx 1-2$. This suggests that
models with strong evolution ($\zeta > 2$) are not favoured by the
observed non-evolving luminosity function. On the other hand, there
must be some \emph{mild} evolution in the $L-M$ relations ($\zeta
\gsim 1$) in order to reproduce the high redshift luminosity
function. This reasoning holds for a gaussian primordial density
field.  In any case, if one finds  using direct measurements of the
mass and luminosity that $\zeta < 1$, it would then be difficult to
explain the non-evolving luminosity function within the gaussian
paradigm on cluster scales.

\subsubsection{Estimating the unresolved thermal SZ contribution} 
\label{sec:Simus:Compare:CMB}

\begin{figure}
\epsfig{file=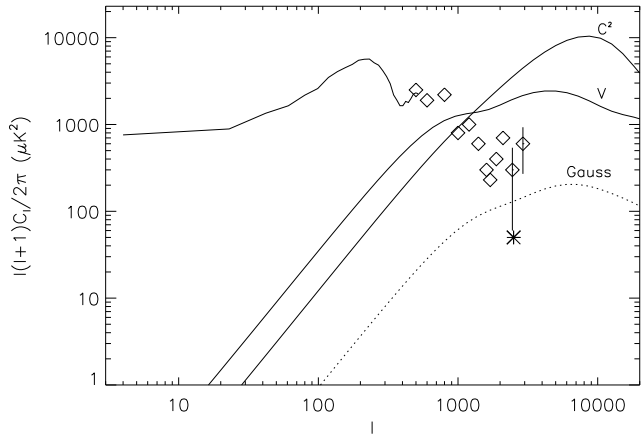,width=9cm}
\caption{\emph{WMAP} temperature angular power spectrum (upper left
curve) compared with the cluster thermal SZ power spectrum computed in
non-gaussian simulations (solid lines for $V$ and $C^2$) and 
gaussian (dotted line) models for clusters. The $M-T$ relation assumes here $\psi = 1$ and the
$\beta$-model $r_{\tx{core}}$=0.08 \hMpcDot The diamonds and vertical
lines report the observations and the star is the extrapolation of the
primary CMB power spectrum. Positive evolution in $M-T$ and compact
clusters increase the power at small scales.}
   \label{fig:Cl_ACBAR_CBI}
\end{figure}

\begin{figure}
\epsfig{file=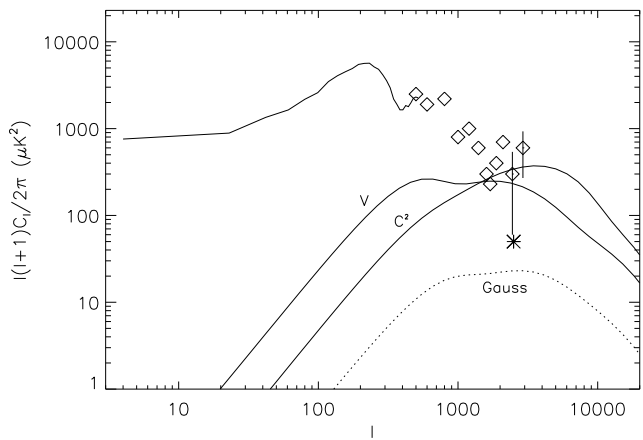,width=9cm}
   \caption{Same as Fig.~\ref{fig:Cl_ACBAR_CBI} but for a $M-T$
relation with $\psi =0$ and a $\beta$-model with $r_{\tx{core}}$=0.15
\hMpcDot}
   \label{fig:Cl_ACBAR_CBI_B}
\end{figure}

Figure~\ref{fig:Cl_ACBAR_CBI} illustrates the sensitivity of the SZ
power spectrum at small angular scales to the abundance of the high
redshift population of clusters.  In practice, the computation of the
cluster SZ thermal power spectrum involves an integral of the mass
function where each mass must be assigned a temperature.  To compute
the components $C_{\ell}$ of the angular power spectrum, we follow the
formalism of \citet{Die03}.  We write the $M-T$ scaling relation as:
\begin{equation}
T \propto M^{\chi} (1+z)^{\psi}.
\label{eq:TM}
\end{equation}
In this paragraph, the gas distribution inside the cluster needs
modelling as well.  For simplicity, we assume a $\beta$-model with
$\beta=2/3$ and core radius $r_{\rmn{core}}$.  We normalise the total
emission to the total mass and the temperature. The angular power
spectrum scales quadratically with the constant in front of
equation~\ref{eq:TM} and we set it so that $T = 8.5$ keV for a $M =
10^{15} h^{-1} M_{\odot}$ cluster at $z=0$. The results depend only
mildly on $\chi$ and we fix it to $\chi = 0.6$.  Our free parameters
are therefore the redshift dependence of $\psi$ and the mass- and
redshift-dependence of the core radius $r_{\rmn{core}}$ which we
parametrise as:
\begin{equation}
r_{\rmn{core}} = r_{0} M^{1/3}/(1+z)
\end{equation}
Fig.~\ref{fig:Cl_ACBAR_CBI} shows the cluster thermal SZ power
spectrum for the gaussian (dotted line) and non-gaussian models (solid
lines), taking $\psi = 1$ and $r_{\tx{core}}$=0.08 \hMpc (for $M =
10^{15} h^{-1} M_{\odot}$ at z=0).  The diamonds 
and vertical lines report the ACBAR and CBI observations, the star is
the extrapolation of the primary CMB power spectrum, and the upper
left curve is a fit to the \emph{WMAP} measurements of the temperature
power spectrum. In this plot, contrary to the previous paragraph, we
did not require   non-gaussian and gaussian models to have exactly the
same abundance of massive clusters at $z=0$, but respect the
normalisation given in Fig.~\ref{fig:MF}.  The non-gaussian
predictions clearly exceed the observations at $\ell\sim2500$, while the
gaussian case underpredicts the power. To see how this depends on the
free parameters of the modelling, Fig.~\ref{fig:Cl_ACBAR_CBI_B}
repeats the exercise for $\psi =0$ and $r_{\tx{core}}$=0.15 \hMpcDot
The effect of changing these parameters is significant (mainly the
increase in $r_{\tx{core}}$): predictions for the non-gaussian models
agree with the observed power while the gaussian case has too little
power with such a set of parameters. At $\ell=2000$, we note that the
thermal SZ power spectrum shown in Fig.~\ref{fig:Cl_ACBAR_CBI_B} for
the gaussian case agrees within a factor 1.5 with that obtained by
\citet{Zha02} using hydrodynamical simulations.

In fact, rather than the  absolute level, it is the relative excess of
the thermal SZ power spectrum in the non-gaussian models compared to
the gaussian case which is interesting.  Typically, at scales
$\ell\gsim2000$ non-gaussian models have more than an order of magnitude
more power than the gaussian case.  If $\sigma_{8}=0.9$ and if the
excess power seen by CBI and ACBAR is mostly due to the cluster
thermal SZ effect, this makes a strong case for non-gaussianity. Even
if the excess factor we find for the power in the non-gaussian case
were too high to fit observations given the local abundance of
clusters, much room would be left for mildly non-gaussian models with
only reduced departure from gaussianity (the $V$ and $C^2$ simulations
used here have significant non-gaussianity).

\section[]{Conclusions}
\label{sec:CCL}

While CMB-scale observations have started to constrain the possible
amounts of non-gaussianity in the primordial cosmic density field,
models with scale-dependent non-gaussianity could remain attractive
for several aspects of structure formation if they predict 
significant non-gaussianity on cluster scales.  Here, we have addressed  
whether such non-gaussianity with reasonable scaling relations can
indeed explain the observables at high and low redshifts.

We have confronted three recent observational hints of early cluster
formation and reduced late-time X-ray cluster abundance evolution to
the predictions of two phenomenological models with strong
non-gaussianity on galaxy to supercluster scales and opposite
amounts of primordial skewness.  Our approach using collisionless simulations was
mainly illustrative; for instance, we did not fit the parameters of
the models we employed so that they precisely match the present-day
galaxy power spectrum, but we have simply normalised the models so
that they approximately reproduce the abundance of local clusters
predicted in the concordance \lcdm cosmology.

Rather, we probed whether strongly non-gaussian models such as a model
with primordial voids with negatively skewed overdensity distribution,
and a $\chi$-square model with one degree of freedom with positively
skewed overdensity distribution would help resolve three possible
difficulties faced by the gaussian \lcdm paradigm.

Firstly, we have found the  protocluster observed by
\citet{Mil04} to be significantly easier to form in the two
non-gaussian models we have simulated than in the \lcdm case, where
its estimated velocity dispersion seems difficult to reach.

Secondly, we have compared the slow bright X-ray cluster abundance
evolution observed in the 160 deg$^2$ \emph{ROSAT} survey at
$z\lsim0.6$ to the predictions of the non-gaussian models using two
models for the redshift evolution of the $L-M$ relation. To reproduce
the data, the simulated gaussian model required significant $L-M$
evolution, which is observationally disfavoured.  In the case of the
two non-gaussian models, both evolving and non-evolving $L-M$
relations provided predictions bracketing the data.

Thirdly, we tackled the excess of power in CMB temperature maps seen
by the interferometers at $\ell\sim 2500$. Using a simple $\beta$-model
for the intracluster gas distribution, we found that the thermal SZ
contribution to the CMB temperature power spectrum is typically an
order of magnitude higher in the non-gaussian models than in the
gaussian case, irrespective of the two normalisations we have tried
for the core radius of the intracluster gas and for the evolution of
the $T-M$ relation. If the observed excess of power is due to the
thermal SZ effect, non-gaussianity may be required for \lcdm to also
match the local abundance of clusters (an alternative being an adjustment 
of the scaling relations).  In fact, more ``moderate'' amounts of 
non-gaussianity than the ones considered here might suffice, and 
the room between the two strongly non-gaussian models we have 
simulated and a gaussian primordial density field should be explored in future work.

To summarise, we have found that cluster-scale non-gaussianity can
simultaneously explain an important constraint at high redshift, and
two observables at lower redshift when we use proper scaling
relations.

New constraints on  possible non-gaussianity of the primordial
density field on relatively small scales will come from large weak
lensing surveys.  They are expected by \citet{Am03} to yield
constraints at a level similar to those set by current all-sky CMB
maps on very large scales. In addition,  both the inner density profiles of large DM haloes and the 
abundance of substructure are sensitive to the sign of the skewness 
of the primordial graviational potential field \citep{Av03}.  Finally, 
new measurements of the 2 and n-point 
correlation function of local galaxy clusters of different masses will
also set tight constraints. If they yield negligible cluster scale
non-gaussianity then cluster formation and evolution might need
substantial revisions. If not, simulations such as those presented
here will need to be generalised.

\section{Acknowledgements}

We thank P. Avelino, V. Avila-Reese and S. Majumdar for comments. 
HM thanks PPARC for support. JMD acknowledges 
support by the David and Lucile Packard Foundation 
and the Cottrell foundation. 

\bsp \bibliographystyle{mnras}

\begin{thebibliography}{44}
\expandafter\ifx\csname natexlab\endcsname\relax\def\natexlab#1{#1}\fi

\bibitem[Amara \& Refregier(2003)]{Am03}
Amara A., Refregier A., 2003, preprint, astro-ph/0310345

\bibitem[Avelino \& Liddle(2004)]{Ave04}
Avelino P.~P., Liddle A.~R., 2004, MNRAS, 348, 105

\bibitem[Avelino et~al.(1998)Avelino, Shellard, Wu \& Allen]{Ave98}
Avelino P.~P., Shellard E. P.~S., Wu J. H.~P., Allen B., 1998, ApJ, 507, 101

\bibitem[{Avila-Reese} et~al.(2003){Avila-Reese}, {Col{\'{\i}}n}, {Piccinelli}
  \& {Firmani}]{Av03}
{Avila-Reese} V., {Col{\'{\i}}n} P., {Piccinelli} G., {Firmani} C., 2003, ApJ,
  598, 36

\bibitem[Bahcall et~al.(2003)Bahcall, Dong, Bode et~al.]{Bah03a}
Bahcall N.~A., Dong F., Bode P., et~al., 2003, ApJ, 585, 182

\bibitem[{Borgani} et~al.(2001){Borgani}, {Rosati}, {Tozzi} et~al.]{Bor01}
{Borgani} S., {Rosati} P., {Tozzi} P., et~al., 2001, ApJ, 561, 13

\bibitem[Broadhurst et~al.(1990)Broadhurst, Ellis, Koo \& Szalay]{Broad90}
Broadhurst T.~J., Ellis R.~S., Koo D.~C., Szalay A.~S., 1990, Nature, 343, 726

\bibitem[{Chen} et~al.(2003){Chen}, {Cooray}, {Yoshida} \& {Sugiyama}]{Che03}
{Chen} X., {Cooray} A., {Yoshida} N., {Sugiyama} N., 2003, MNRAS, 346, L31

\bibitem[{Diego} et~al.(2003){Diego}, {Silk} \& {Sliwa}]{Die03}
{Diego} J.~M., {Silk} J., {Sliwa} W., 2003, MNRAS, 346, 940

\bibitem[El-Ad \& Piran(2000)]{ElAd00}
El-Ad H., Piran T., 2000, MNRAS, 313, 553

\bibitem[{Frith} et~al.(2003){Frith}, {Busswell}, {Fong}, {Metcalfe} \&
  {Shanks}]{Fri03}
{Frith} W.~J., {Busswell} G.~S., {Fong} R., {Metcalfe} N., {Shanks} T., 2003,
  MNRAS, 345, 1049

\bibitem[Griffiths et~al.(2003)Griffiths, Kunz \& Silk]{Gri03}
Griffiths L.~M., Kunz M., Silk J., 2003, MNRAS, 339, 680

\bibitem[Hoekstra et~al.(2000)Hoekstra, Franx \& Kuijken]{Hoe00}
Hoekstra H., Franx M., Kuijken K., 2000, ApJ, 532, 88

\bibitem[{Holden} et~al.(2002){Holden}, {Stanford}, {Squires} et~al.]{Hold02}
{Holden} B.~P., {Stanford} S.~A., {Squires} G.~K., et~al., 2002, AJ, 124, 33

\bibitem[Hoyle \& Vogeley(2003)]{Hoy03}
Hoyle F., Vogeley M.~S., 2003, preprint, astro-ph/0312533

\bibitem[Komatsu et~al.(2003)Komatsu, Kogut, Nolta et~al.]{Kom03}
Komatsu E., Kogut A., Nolta M.~R., et~al., 2003, preprint, astro-ph/0302223

\bibitem[Koyama et~al.(1999)Koyama, Soda \& Taruya]{Koy99}
Koyama K., Soda J., Taruya A., 1999, MNRAS, 310, 1111

\bibitem[{Kuo} et~al.(2004){Kuo}, {Ade}, {Bock} et~al.]{Kuo04}
{Kuo} C.~L., {Ade} P.~A.~R., {Bock} J.~J., et~al., 2004, ApJ, 600, 32

\bibitem[Kurk et~al.(2003)Kurk, Venemans, Rottgering, Miley \&
  Pentericci]{Kur03}
Kurk J., Venemans B., Rottgering H., Miley G., Pentericci L., 2003, preprint,
  astro-ph/0309675

\bibitem[{Majumdar}(2001)]{Maj01}
{Majumdar} S., 2001, ApJ, 555, L7

\bibitem[{Majumdar} \& {Subrahmanyan}(2000)]{Maj00}
{Majumdar} S., {Subrahmanyan} R., 2000, MNRAS, 312, 724

\bibitem[Mason et~al.(2003)Mason, Pearson, Readhead et~al.]{Mas03}
Mason B., Pearson T.~J., Readhead A. C.~S., et~al., 2003, ApJ, 591, 540

\bibitem[Matarrese et~al.(2000)Matarrese, Verde \& Jimenez]{Mat00}
Matarrese S., Verde L., Jimenez R., 2000, ApJ, 541, 10

\bibitem[Mathis et~al.(2003)Mathis, Silk, Griffiths \& Kunz]{Ma03a}
Mathis H., Silk J., Griffiths L.~M., Kunz M., 2003, preprint, astro-ph/0303519

\bibitem[Miley et~al.(2004)Miley, Overzier, Tsvetanov \& {the ACS/GTO
  Team}]{Mil04}
Miley G.~K., Overzier R.~A., Tsvetanov Z.~I., {the ACS/GTO Team}, 2004,
  preprint, astro-ph/0401034

\bibitem[{Moretti} et~al.(2001){Moretti}, {Guzzo}, {Campana} et~al.]{Mor01}
{Moretti} A., {Guzzo} L., {Campana} S., et~al., 2001, in { ASP Conf. Ser. 234:
  X-ray Astronomy 2000\/},  393--+

\bibitem[{Mullis} et~al.(2003){Mullis}, {McNamara}, {Quintana} et~al.]{Mul02}
{Mullis} C.~R., {McNamara} B.~R., {Quintana} H., et~al., 2003, ApJ, 594, 154

\bibitem[Mullis et~al.(2004)Mullis, Vikhlinin, Henry et~al.]{Mul04}
Mullis C.~R., Vikhlinin A., Henry J.~P., et~al., 2004, preprint,
  astro-ph/0401605

\bibitem[Neumann \& Arnaud(2000)]{Neu00}
Neumann D.~M., Arnaud M., 2000, ApJ, 542, 35

\bibitem[Peebles(1999)]{Pee99a}
Peebles P. J.~E., 1999, ApJ, 510, 523

\bibitem[Peebles(2001)]{Pee01}
Peebles P. J.~E., 2001, ApJ, 557, 495

\bibitem[Robinson \& Baker(2000)]{RB00}
Robinson J., Baker J.~E., 2000, MNRAS, 311, 781

\bibitem[Robinson et~al.(2000)Robinson, Gawiser \& Silk]{RGS00}
Robinson J., Gawiser E., Silk J., 2000, ApJ, 532, 1

\bibitem[{Rosati} et~al.(2002){Rosati}, {Borgani} \& {Norman}]{Ros02}
{Rosati} P., {Borgani} S., {Norman} C., 2002, ARA\&A, 40, 539

\bibitem[{Rosati} et~al.(2000){Rosati}, {Giacconi}, {Bergeron} et~al.]{Ros00}
{Rosati} P., {Giacconi} R., {Bergeron} J., et~al., 2000, Bulletin of the
  American Astronomical Society, 32, 726

\bibitem[{Sadat} et~al.(1998){Sadat}, {Blanchard} \& {Oukbir}]{Sad98}
{Sadat} R., {Blanchard} A., {Oukbir} J., 1998, AAP, 329, 21

\bibitem[Vauclair et~al.(2003)Vauclair, Blanchard, Sadat et~al.]{Vau03}
Vauclair S.~C., Blanchard A., Sadat R., et~al., 2003, A\&A, 412, 37

\bibitem[{Vikhlinin} et~al.(1998){Vikhlinin}, {McNamara}, {Forman}, {Jones},
  {Quintana} \& {Hornstrup}]{Vik98}
{Vikhlinin} A., {McNamara} B.~R., {Forman} W., {Jones} C., {Quintana} H.,
  {Hornstrup} A., 1998, ApJ, 498, L21+

\bibitem[{Vikhlinin} et~al.(2002){Vikhlinin}, {VanSpeybroeck}, {Markevitch},
  {Forman} \& {Grego}]{Vik02b}
{Vikhlinin} A., {VanSpeybroeck} L., {Markevitch} M., {Forman} W.~R., {Grego}
  L., 2002, ApJ, 578, L107

\bibitem[Vikhlinin et~al.(2003)Vikhlinin, Voevodkin, Mullis et~al.]{Vik03}
Vikhlinin A., Voevodkin A., Mullis C.~R., et~al., 2003, ApJ, 590, 15

\bibitem[Voevodkin \& Vikhlinin(2004)]{Voe04}
Voevodkin A., Vikhlinin A., 2004, ApJ, 601, 610

\bibitem[Weinberg \& Cole(1992)]{Wein92}
Weinberg D.~H., Cole S., 1992, MNRAS, 259, 652

\bibitem[Willick(2000)]{Wi00}
Willick J.~A., 2000, ApJ, 530, 780

\bibitem[{Zhang} et~al.(2002){Zhang}, {Pen} \& {Wang}]{Zha02}
{Zhang} P., {Pen} U., {Wang} B., 2002, ApJ, 577, 555

\end{thebibliography}

\end{document}